\documentclass[9pt,preprint]{aastex}

\usepackage{spr-astr-addons}
\usepackage{graphicx}
\usepackage{amssymb}
\usepackage{amsmath}
\usepackage{url}

\newcommand{\E}[1]{\times 10^{#1}}
\newcommand{\dee}{{\rm d}}
\newcommand{\m}{\rm m}

\newcommand{\unit}[1]{\;\textrm{#1}}

\begin{document}

\title{Scalar field power-law cosmology with spatial curvature and dark energy-dark matter interaction}

\author{Burin Gumjudpai\altaffilmark{1,2}} 
\email{buring@nu.ac.th}
\and
\author{Kiattisak Thepsuriya\altaffilmark{1}}
\email{kiattisak@physics.org}

\altaffiltext{1}{ThEP's CRL, NEP, The Institute for Fundamental Study, Naresuan University, Phitsanulok 65000, Thailand}
\altaffiltext{2}{Thailand Center of Excellence in Physics, Ministry of Education, Bangkok 10400, Thailand}

\begin{abstract}
We consider a late closed universe of which scale factor is a power function of time using observational data from combined WMAP5+BAO+SNIa dataset and WMAP5 dataset. The WMAP5 data give power-law exponent, $\alpha = 1.01$ agreeing with the previous study of $H(z)$ data while combined data gives $\alpha=0.985$. Considering a scalar field dark energy and dust fluid evolving in the power-law universe, we find field potential, field solution and equation of state parameters. Decaying from dark matter into dark energy is allowed in addition to the non-interaction case. Time scale characterizing domination of the kinematic expansion terms over the dust and curvature terms in the scalar field potential are found to be approximately 5.3 to 5.5 Gyr. The interaction affects in slightly lowering the height of scalar potential and slightly shifting potential curves rightwards to later time. Mass potential function of the interacting Lagrangian term is found to be exponentially decay function.

\end{abstract}



\maketitle

\section{Introduction}
The presence of a scalar field is motivated by ideas in high energy physics and quantum gravities, although it has not been discovered experimentally. TeV-scale experiments at LHC and Tevatron might be able to confirm its existence. It is nevertheless widely accepted in several theoretical modeling frameworks, especially in contemporary cosmology of which an early-time accelerating expansion, the inflation is proposed to be driven by a scalar field in order to solve horizon and flatness problems \citep{star80}. After inflation, components of barotropic fluids such as radiation and other non-relativistic matter were produced during reheating and cooling-down processes. The present universe is also found to be in acceleration phase which is strongly backed up by various observations, e.g. the cosmic microwave background \citep{Masi:2002hp}, large-scale structure surveys \citep{Scranton:2003in} and supernovae (SN) type Ia observations \citep{Riess:1998cb, Riess:2004nr, Astier:2005qq}. The scalar field could be responsible to the present acceleration in various models of dark energy \citep{padma04}. 

Power-law cosmology with $a \propto t^{\alpha}$, describes an acceleration phase for with $ \alpha > 1$. The model provides simple model of expansion as well as a convenient way to fit with observations. Although tightly constrained by nucleosynthesis \citep{Sethi:1999sq, Kaplinghat:1999}, the power-law expansion with $\alpha \sim 1$, could be considered viable at later time sufficiently long after matter-radiation equality era.  Studies of high-redshift objects such as globular clusters \citep{Kaplinghat:1999, Lohiya:1997ti, Dev:2002sz, Sethi2005}, SN Ia data \citep{Sethi2005, Kumar2011}, SN Ia with $H(z)$ data \citep{Dev2001, Sethi2005, Dev:2008ey} and of X-ray gas mass fraction measurement of galaxy clusters \citep{Allen2002, Zhu:2007tm} within context of power-law cosmology have been reported from time to time. Moreover other aspects such as gravitational lensing statistics \citep{Dev:2002sz} and angular size-redshift data of compact radio sources \citep{Jain2003} of the power-law cosmology can be applied to constrain the model.  Motivations of the model are various. Simplest inflationary model with power-law expansion can remove flatness and horizon problems leaving with simple spectrum \citep{Lucchin}. In present universe, idea of linear coasting cosmology ($a \propto t$) \citep{Kolb1989} are proposed such that it can resolve age problem in the CDM model \citep{Lohiya:1997ti}. On fundamental physics motivation, the linear coasting model can arise from non-minimally coupled scalar-tensor theory in which the scalar field couples to the curvature contributing to energy density that cancels out vacuum energy \citep{Sethi:1999sq, Ford1987}. The model could also be a result of domination of an SU(2) cosmological instanton in the early universe \citep{Allen1999}.

We consider a non-flat ($k \neq 0$) FLRW universe with expansion in the form of the power law function. The power-law expansion occurs well after matter-radiation equality era. Two major ingredients are scalar field dark energy evolving under the scalar field potential $V(\phi)$, and barotropic (dust) fluid consisting of cold dark matter and baryons. We use combined WMAP5+BAO+SN data and the WMAP5 data alone  \citep{Hinshaw:2008kr} to determine scalar field potential, values of $\alpha$ and other relevant parameters. Other related study for scalar field power-law cosmology are for the flat case and with phantom field \citep{Kaeonikhom:2010vq}.
 Previously, not within the context of power-law cosmology, there are a number of works in constructing scalar field potential directly from observational data or in theoretical construction of potential under specific assumption. These are considered in various situations, for example, construction of the potential using SNe Ia dataset \citep{simon05}, construction of potential assuming scaling solution \citep{astro-ph/9809272}, case of scalar field negative potential without using observational data \citep{Cardenas:2004ji}, construction of scalar field potential using dark energy density function with non-specified expansion law \citep{Guo:2006ab} and other several works in \citep{Hut:1999}. In this work, we also consider situation when there is interaction between cold dark matter and dark energy with constant interaction rate \citep{Amendola:1999er}. We see how the interaction can affect scalar field potential and we will find mass potential function of the interaction Lagrangian. We stress that the WMAP5 combined dataset allows non-flat universe and dark energy equation of state. This is because the addition of  the SN Ia analysis to  the WMAP5 dataset marginalize the equation of state parameter and spatial curvature at the same time. The later WMAP7 dataset is not suitable our study since its relevant derived parameters are found under flat assumption, allowing only studying of equation of state.  

\section{Scalar-field power-law cosmology}

Here we will work with observational data in SI units. Two perfect fluids, the dust matter and scalar field $\phi \equiv \phi(t)$, in the late FLRW universe of  $w$CDM model with zero cosmological constant are considered. The time evolution of the barotropic fluid  (cold dark matter and baryonic matter) is governed by the fluid equation
\begin{equation}
\dot{\rho}_{\m} = -3H\rho_{\m},
\end{equation}
since $w_{\m}$ is constant,
\begin{equation}\label{energydensity} \rho_{\m} = \frac{D}{a^n}, \end{equation}
where $n \equiv 3 (1 + w_{\m})$ and $D \geq 0$ is a proportional constant. For the scalar field, supposed that it is minimally coupled to gravity, its Lagrangian density is $\mathcal{L} = \dot{\phi}^2 /2 - V(\phi)$.  The energy density and pressure are
\begin{align}   \label{eqEoS}
\rho_\phi c^2 = \frac{1}{2} \dot{\phi}^2 + V(\phi),\quad p_\phi = \frac{1}{2} \dot{\phi}^2 - V(\phi).
\end{align}
The fluid equation of the field describing its energy conservation as the universe expands is
\begin{equation} 
\ddot{\phi} + 3H\dot{\phi} + \frac{\dee}{\dee \phi}V = 0.  \label{KG}
\end{equation}
Total energy density $\rho_\mathrm{tot}$ and total pressure $p_\mathrm{tot}$ of the mixture are simply the sums of those contributed by each fluid, for which the Friedmann equation in SI unit is
\begin{equation}
H^2 = \frac{8 \pi G}{3} \rho_\mathrm{tot} - \frac{k c^2}{a^2}.
\end{equation}
It is straightforward to show that
\begin{equation}\label{vraw} 
V = \frac{3 M_{\rm P}^2 c}{\hbar } \left( H^2 + \frac{\dot{H}}{3} + \frac{2k c^2}{3a^2} \right) + \left( \frac{n - 6}{6} \right) \frac{D c^2}{a^n}\,.
\end{equation}
The reduced Planck mass, $M_{\rm P}$ is related to Newton's constant $G$ as $M_{\rm P}^2 = \hbar c/8 \pi G$. For dust fluid $n = 3$ ($w_{\m} = 0$), the last term is just 
$-{Dc^2}/{2a^3}$. In power-law cosmology, scale factor is a power function of cosmic time
\begin{equation}\label{scalefactor} 
a(t) = a_0 \left( \frac{t}{t_0} \right)^{\alpha}\,, 
\end{equation}
Hence the Hubble parameter is
\begin{equation}\label{hubbleparameter} H(t) = \frac{\dot{a}(t)}{a(t)} = \frac{\alpha}{t}. \end{equation}
 Incorporating \eqref{scalefactor} and \eqref{hubbleparameter} into the above equation, we obtain
\begin{equation}\label{vt}
V(t) = \frac{M_\mathrm{P}^2 c}{\hbar} \left( \frac{3{\alpha}^2 - {\alpha}}{t^2} + \frac{2kc^2 t_0^{2{\alpha}}}{a_0^2 t^{2{\alpha}}} \right) - \frac{Dc^2}{2}\frac{t_0^{3{\alpha}}}{a_0^3 t^{3{\alpha}}}.
\end{equation}
We see that $\lim_{t \to \infty} V(t) = 0$ at late time. We will consider contribution of the first term (effects of acceleration rate and Hubble expansion) alone in comparison to total contribution when including the second (the curvature) and the third (dust density) terms.

\section{Cosmological parameters and observational data}

Using  \eqref{hubbleparameter} at the present time, we have
$ \alpha = H_0 t_0 $.   The sign of $k$ depends on the sign of the density parameter $\Omega_k \equiv -kc^2 / a^2 H^2$. In our convention here, $k = 1 \;(\Omega_{k} < 0)$ for a closed universe, $k = 0$ for a flat one, and $k = -1 \;(\Omega_{k} > 0)$ for an open one. Allowing $k$ to take these three discrete values $(0, \pm 1)$, present value of the scale factor, $a_0$ is not fixed to unity and it can be found from the definition of $\Omega_{k,0}$ directly\footnote{When considering case of dark-mater and dark-energy interaction, we set $a_0 = 1$ for convenience.}, 
\begin{equation}
 a_0 = \frac{c}{H_0} \sqrt{\frac{-k}{\Omega_{k, 0}}}. 
 \end{equation}
The density constant $D$ can be found from \eqref{energydensity},
 \begin{equation} 
 D = \rho_{\m, 0}\, a_0^3 = \Omega_{\m, 0}\, \rho_{c, 0} \, a_0^3, 
 \end{equation}
where $ \Omega_{\m, 0} = \Omega_{\mathrm{CDM}, 0} + \Omega_{b, 0},$ i.e. the sum of the present density parameters of the barotropic fluid components. $\rho_{c, 0}$ is  present value of critical density. 
Radiation and neutrino contributions are negligible. The values of $H_0$, $t_0$, $\Omega_{k, 0}, \Omega_{\mathrm{CDM}, 0}$, and $\Omega_{b, 0}$ are taken from observational data. 
Note that although $t_0$ value in the WMAP5 data is derived in $\Lambda$CDM model. This fact can be relaxed and we can estimably use it. This is because the observed equation of state at present is very closed to -1 and $t_0$ can not be derived if neither the equation of state nor $\alpha$ is given. 
  Two sets of data used here are provided by \citep{Hinshaw:2008kr}. One comes solely from the WMAP5 data and the other is the WMAP5 data combined with distance measurements from Type Ia supernovae (SN) and the Baryon Acoustic Oscillations (BAO) in the distribution of galaxies. For $t_0$, $H_0$, $\Omega_{b, 0}$, and $\Omega_{\mathrm{CDM}, 0}$, we take their maximum likelihood values. The curvature density parameter $\Omega_{k, 0}$ comes as a range with 95\% confidence level on deviation from the simplest $\Lambda$CDM model. These data can be seen in Table  \ref{datatable} in which $\Omega_{k, 0}$ is reported as an interval of minimum and maximum values. We use the mean of each $\Omega_{k, 0}$ interval to represent $\Omega_{k, 0}$.
Table \ref{resulttable} presents parameters derived from data in Table \ref{datatable}. The exponent $\alpha$ can be simply found using $ \alpha = H_0 t_0 $.  The WMAP5 data can give an accelerating expansion as $\alpha > 1$ while the combined WMAP5+BAO+SN does not.  The determination of $\alpha$ from other studies are such as X-Ray gas mass fractions in galaxy clusters which favors open universe with $\alpha > 1$ ($\alpha = 1.14 \pm 0.05$) \citep{Zhu:2007tm}.  Larger number of data points are needed in this study in order to distinct cosmic geometry in power-law scenario. Combined analysis from SNLS and $H(z)$ data (from Germini Deep Deep Survey+archival data points)  yields $\alpha = 1.31$ (assuming open geometry) \citep{Dev:2008ey}. Flat and closed geometry are still viable in these studies. Compared to \citep{Dev:2008ey}, without assuming geometry, $H(z)$ (spatial-geometry independent) data alone gives $1.07$ while WMAP5 data alone yields $\alpha = 1.01$. 
Combined $H(z)+$SN Ia data (SNLS) assuming closed geometry gives $\alpha = 2.28$ \citep{Dev:2008ey}. Although 
Table  \ref{resulttable} shows mean values of $\bar{\Omega}_{k, 0}$ in closed geometry region, but it is clear that with closed geometry, our values of $\alpha$ do not match the $\alpha$ value from $H(z)+$SN Ia data. The curvature dependence in SN data analysis could give larger deviation in $\alpha$ from unity. Better SN data analysis would improve determination of spatial curvature in power-law cosmology.

\begin{table*} 
\caption{Observational derived parameters from \citep{Hinshaw:2008kr}.}
\begin{tabular}{ccc}
\tableline
\textbf{Parameters} & \textbf{WMAP5+BAO+SN} & \textbf{WMAP5}\\
\tableline
$t_0$ & $13.72$ Gyr & $13.69$ Gyr\\
$H_0$ & $70.2$ km/s/Mpc & $72.4$ km/s/Mpc\\
$\Omega_{b, 0}$ & $0.0459$ & $0.0432$\\
$\Omega_{\mathrm{CDM}, 0}$ & $0.231$ & $0.206$\\
$\Omega_{k, 0}$ & $-0.0179 < \Omega_{k, 0} < 0.0081$ & $-0.063 < \Omega_{k, 0} < 0.017$\\
$\rho_{c, 0}$ & $9.26 \times 10^{-27} $ kg$/$m$^3$  & $  9.85 \times 10^{-27} $   kg$/$m$^3$ \\
$\rho_{b, 0}$ & $ 4.25 \times 10^{-28} $  kg$/$m$^3$ & $ 4.25 \times 10^{-28} $   kg$/$m$^3$ \\
$\rho_{\rm m, 0}$ & $ 2.56 \times 10^{-27} $  kg$/$m$^3$   & $ 2.45 \times 10^{-27} $   kg$/$m$^3$  \\
\tableline
\end{tabular} \label{datatable}
\end{table*}

\begin{table*}
\caption{A summary of result for scalar field power-law cosmology. Times are shown in Gyr for comprehensibility. Positive and negative $\Omega_k$'s correspond to open and closed universes, respectively. For the interaction case, $a_0$ is set to 1.}
\begin{tabular}{ccccc}
\tableline \hline
& \multicolumn{2}{c}{\textbf{WMAP5+BAO+SN}} & \multicolumn{2}{c}{\textbf{WMAP5}}\\
\tableline
$\alpha$ & \multicolumn{2}{c} {0.985}    & \multicolumn{2}{c}{1.01}  \\  \tableline  
& $\bar{\Omega}_{k, 0} = -0.0049$ & $-0.0179 < \Omega_{k, 0} < 0.0081$ & $\bar{\Omega}_{k, 0} = -0.023$ & $-0.063 < \Omega_{k, 0} < 0.017$\\
\hline
{$a_0$ (non-int. case)}  & {$1.9\E{27}$} & $a_0 > 9.85\E{26}$ (closed) & {$8.4\E{26}$} & $a_0 >5.1\E{26}$ (closed)\\ 
   &   & $a_0 > 1.5\E{27}$ (open) && $a_0 > 9.8\E{26}$ (open)\\  
 \tableline
 non-interaction case   &   &    &    & \\ 
$t_\textrm{intercept}$  & $2.65$ Gyr & $2.62$ Gyr $< t < 2.69$ Gyr & $2.65$ Gyr & $2.55$ Gyr $< t < 2.76$ Gyr\\
$t_\textrm{max} $  & $3.99$ Gyr & $3.94$ Gyr $< t < 4.01$ Gyr & $3.97$ Gyr & $3.82$ Gyr $< t < 4.13$ Gyr\\
$t_\textrm{inflection}$  & $5.33$ Gyr & $5.27$ Gyr $< t < 5.40$ Gyr & $5.29$ Gyr & $5.09$ Gyr $< t < 5.50$ Gyr\\ \tableline
interaction case &          &         &   \\
$t_\textrm{intercept}$  & $2.77$ Gyr & $2.73$ Gyr $< t < 2.80$ Gyr & $2.76$ Gyr & $2.66$ Gyr $< t < 2.87$ Gyr\\
$t_\textrm{max} $  & $4.16$ Gyr & $4.10$ Gyr $< t < 4.21 $ Gyr & $4.12$ Gyr & $3.97$ Gyr $< t < 4.28$ Gyr\\
$t_\textrm{inflection}$  & $5.55$ Gyr & $5.48$ Gyr $< t <  5.62 $ Gyr & $5.48$ Gyr & $5.28$ Gyr $< t < 5.70$ Gyr\\
\tableline
\end{tabular} \label{resulttable}
\end{table*}

\section{Results : non-interacting case}

\begin{figure}
\centering
\includegraphics[width=3.3in]{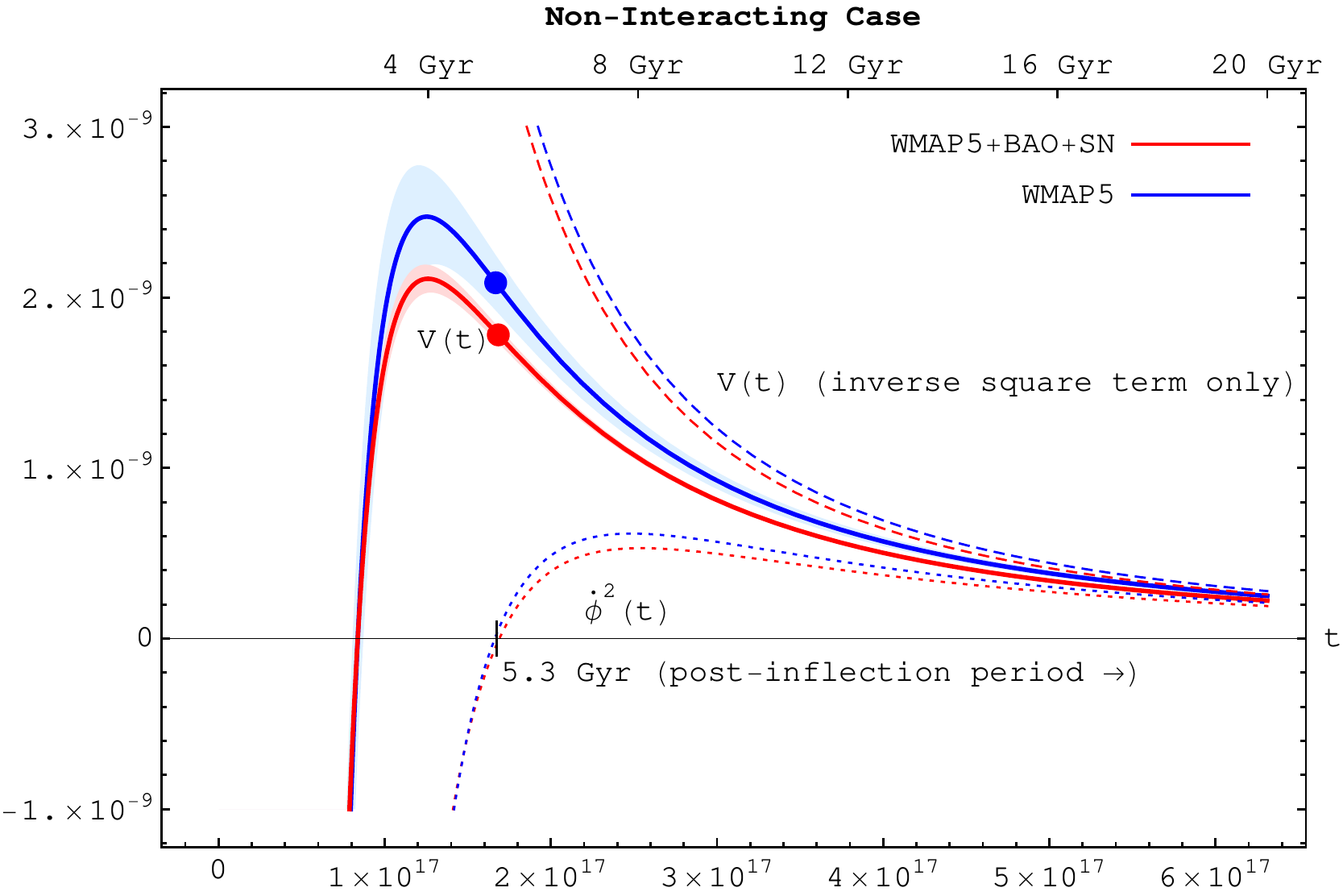} 
\includegraphics[width=3.3in]{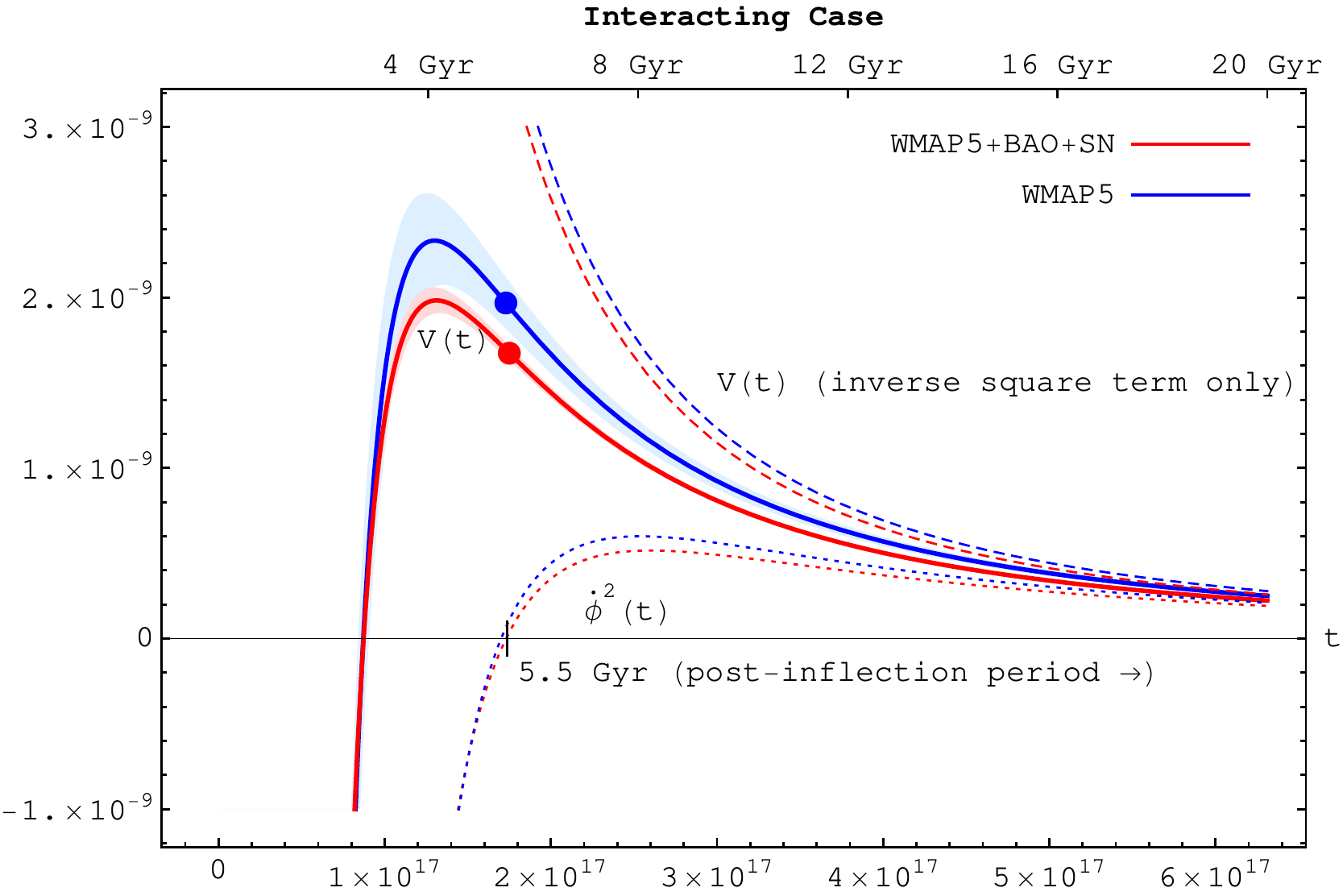}
\caption{Potential plots for non-interaction case (upper panel) and interaction case (lower panel). The potentials are shown in solid lines. The units of the horizontal and vertical axes are sec (or Gyr) and J$/$m$^3$, respectively. Dots mark their inflection points.   The light-shaded regions correspond to ranges of $\Omega_{k, 0}$, and the central solid lines plotted using $\bar\Omega_{k, 0}$. 
  Dashed lines are the potentials' first terms, showing that they are dominant at late time. Plotted in dotted lines are $\dot\phi^2 (t)$ which is positive after  $t_\textrm{inflection}$ 
 afterwhich $\phi (t)$ becomes real. In addition, each potential does not actually converge to its first term, but later intersect with and deviate from it, although they are still very close together (not shown). However, this occurs much later (at $t = 8.8\E{67} \unit{Gyr}$ in both cases). The potential is lower in presence of the interaction between dark matter and dark energy.}
\label{potentialplots}
\end{figure}

Using combined WMAP5+BAO+SN dataset, the potential is
\begin{equation}\label{firstpotential} 
V(t) = \frac{1.03\E{26}}{t^2} + \frac{1.51\E{23}}{t^{1.97}} - \frac{1.50\E{42}}{t^{2.95}}, \end{equation}
whereas, for WMAP5 dataset alone,
\begin{equation}\label{secondpotential}
 V(t) = \frac{1.11\E{26}}{t^2} + \frac{7.67\E{24}}{t^{2.03}} - \frac{4.69\E{43}}{t^{3.04}}. 
 \end{equation}
in SI units.  Their plots are shown in Fig. \ref{potentialplots}. The points at which the potential, its derivative, and its second-order derivative, are zero ($t_\textrm{intercept}$, $t_\textrm{max}$, and $t_\textrm{inflection}$, respectively) are also determined, for both $\bar{\Omega}_{k, 0}$ and each end of the $\Omega_{k, 0}$ interval. The results are summarized in Table \ref{resulttable}.
After $t_\textrm{inflection}$, the potential from each data behaves like its first term, i.e. decreasing in its value while increasing in its slope (being less and less negative). The other terms quickly become weaker. This can be seen in Fig. \ref{potentialplots}. Since the first term is contributed only by $H(t)$ (and its time derivative), it is dominant in the post-inflection phase. In fact, the convergence to zero of the potential is slower than its first term alone (see (\ref{firstpotential}) and (\ref{secondpotential})), because the sum of the last two terms consequently becomes positive before converging to zero. This means that the plots of each potential and its first term in Fig. \ref{potentialplots} eventually crosses, but it occurs much, much later at $t = 8.8\E{67}$ Gyr. Along with the scalar potential, we also obtain
\begin{equation} 
\dot{\phi}^2(t) = - \frac{2 M_\mathrm{P}^2 c}{\hbar} \left( \dot{H} - \frac{kc^2}{a^2} \right) - \frac{D c^2}{a^3} \,,  \label{phidots} 
\end{equation}
in SI units. Using WMAP5+BAO+SN dataset and positive root,
\begin{equation}\label{firstscalarfield}
 \phi(t) = \int_t^{t_0} {\Bigg{(}}\frac{1.05\cdot 10^{26}}{{\tilde t}^2} + \frac{1.51\cdot 10^{23}}{{\tilde t}^{1.97}} - \frac{3.01\cdot 10^{42}}{{\tilde t}^{2.95}}{\Bigg{)}}^{1/2} {\dee} {\tilde t}, \end{equation}
where, for WMAP5 dataset alone,
\begin{equation}\label{secondscalarfield}
 \phi(t) = \int_t^{t_0} {\Bigg{(}}\frac{1.09\cdot 10^{26}}{{\tilde t}^2} + \frac{7.67\cdot 10^{24}}{{\tilde t}^{2.03}} - \frac{9.37\cdot 10^{43}}{{\tilde t}^{3.04}}{\Bigg{)}}^{1/2}  {\dee} {\tilde t}. \end{equation}
In the late post-inflection phase, the first term is dominant over the $k$ and $D$ terms then the last two terms of the radicands are negligible (Fig. \ref{potentialplots}). 
The above two equations are approximated to find exact solutions, 
\begin{equation}
\phi \approx \phi_0 - \left({\frac{2 \alpha M_{\rm P}^2 c}{\hbar}} \right)^{1/2}  \, \ln t  
\end{equation}
where $\phi_0 \equiv  \left({{2 \alpha M_{\rm P}^2 c}/{\hbar}} \right)^{1/2} \, \ln t_0 $. Hence numerically using the WMAP5+BAO+SN combined dataset\footnote{Scalar field exact solutions for the power-law cosmology with non-zero curvature and non-zero matter density are reported in \citep{NLS2}}, 
\begin{equation} 
\phi(t) \approx 4.17\E{14} - 1.03\E{13} \ln t  \label{phi103} 
\end{equation}
whereas, for WMAP5 dataset alone,
\begin{equation} 
\phi(t) \approx 4.23\E{14} - 1.04\E{13} \ln t  \label{phi104} 
\end{equation}
The radicand in (\ref{secondscalarfield}) of the WMAP5 dataset is zero at approximately $t_\textrm{inflection} = 5.3$ Gyr, therefore so does $\phi(t)$. While the WMAP5+BAO+SN combined dataset has the zero radicand (hence zero $\phi(t)$) in
(\ref{firstscalarfield}) later at approximately $t = 5.4$ Gyr.  With post-inflection approximation, $t \approx t_0 \exp{(\phi / {\sqrt{2 \alpha M_{\rm P}^2 c/ \hbar}}  )} $, the potential is written as function of the scalar field as 
\begin{eqnarray}
V(\phi) & \approx &  
\frac{M_{\rm P}^2 c}{\hbar} \Bigg[  \left( \frac{3\alpha^2 - \alpha}{t_0^2}  \right) e^{-2 \phi / {\sqrt{2 \alpha M_{\rm P}^2 c/ \hbar}} } \nonumber \\   &   & + \frac{2 k c^2}{a_0^2}e^{-2 \sqrt{\alpha}\phi / \sqrt{2 M_{\rm P}^2 c/ \hbar }}     \Bigg] \nonumber \\ 
&   & -  \frac{D c^2}{2 a_0^3 } e^{-3 \sqrt{\alpha}\phi / \sqrt{2 M_{\rm P}^2 c/ \hbar }}\,.
\end{eqnarray}
Hence, we can plot $V(\phi)$ as
\begin{eqnarray}  
V(\phi) & \approx   &     \, ({2.72\cdot 10^{-12}}){\,e^{1.92\cdot 10^{-13} \phi}}  \nonumber \\ &  & + \, ({5.51\cdot 10^{-10}}){\,e^{1.95\cdot 10^{-13} \phi}}  \nonumber \\
                    &     &  -\, ({1.15\cdot 10^{-10}}){\,e^{2.88\cdot 10^{-13} \phi}}
        \,,
 \end{eqnarray}
for WMAP5+BAO+SN dataset, whereas, for WMAP5 dataset alone,
\begin{eqnarray}
 V(\phi) & \approx &  \,({5.94\cdot 10^{-10}}){\,e^{1.92\cdot 10^{-13} \phi}}   \nonumber \\ &  & + \,({1.36\cdot 10^{24}}){\, e^{-1.94\cdot 10^{-13} \phi}}   \nonumber \\ &  & +  \,({1.11\cdot 10^{26}}){\,e^{-1.92\cdot 10^{-13} \phi}}
         \,. \end{eqnarray}
 The scalar equation of state as function of time can be found directly using (\ref{eqEoS})
 in $w_{\phi} = p_{\phi}/ \rho_{\phi}$, i.e. for WMAP5+BAO+SN  combined data 
 \begin{align}
  w_\phi  = -0.333  &
    + \frac{(4.43 \cdot 10^{18})-7.01 t^{0.955}}{(1.33 \cdot  10^{19})-\left(6.9 \cdot  10^2\right) t^{0.955}- t^{0.985}} \\
    \lim_{t \to \infty} w_\phi (t) =& - 0.333, \;\;\;    w_\phi (t_0) = - 0.446, 
   \end{align}
and for WMAP5 data alone
 \begin{align}
 w_\phi = -0.342 & +\frac{(1.94 \cdot 10^{17})-\left(6.25 \cdot  10^{{-4}}\right) t^{1.01}}{(5.67 \cdot  10^{17})-\left(6.97 \cdot  10^{{-2}}\right) t^{1.01}- t^{1.04}} \\
\lim_{t \to \infty} w_\phi(t) =& - 0.342 \,, \;\;\;\;\;\;\; 
w_\phi (t_0) = - 0.452\,.
\end{align}
The value of equation of state diverges when time satisfies condition,
\begin{equation}
\alpha^2 = \frac{\hbar}{3 M_{\rm P}^2 } \frac{D c t_0^{3\alpha} }{a_0^3 \, t^{3 \alpha -2}}  -   \frac{k c^2 \, t_0^{2 \alpha}}{a_0^2  \, t^{2 \alpha - 2}}\,.
\end{equation}
 We find weighted effective value of equation of state parameter for dust and scalar field densities. That is    
 \begin{equation}
 w_{\rm eff}  =  \frac{p_{\phi}}{\rho_{\rm m} c^2 + \rho_{\phi} c^2}  \,,
 \end{equation}
which diverges when time satisfies 
\begin{equation}
t^{2(\alpha -1)} = - k    \left( \frac{c t_0^{\alpha}}{ \alpha a_0}  \right)^2  \,.
\end{equation}
 Weighted value for WMAP5+BAO+SN  combined data is
 \begin{align}
w_{\rm eff} =  -0.333 & +\frac{-(1.87\cdot  10^4) + 7.01 t^{0.955}}{ (3.01\cdot  10^4) +\left(6.9 \cdot  10^2\right) t^{0.955}+ t^{0.985}},\\
w_{\rm eff} (t_0) = & - 0.323\,,
   \end{align}
and for WMAP5 data alone
\begin{align}
w_{\rm eff} = -0.342 &+\frac{-(1.72 \cdot 10^3)+\left(6.25 \cdot 10^{{-4}}\right) t^{1.01}}{(2.52 \cdot  10^3)+\left(6.97\cdot  10^{{-2}}\right) t^{1.01}+ t^{1.04}} \\
w_{\rm eff}(t_0) =& -0.342\,.
\end{align}
These values do not match WMAP5 observation in $w$CDM model ($w \approx -1$). We will consider interaction case to check if effects from dark matter-dark-energy interaction could alter the equation of state and the scalar potential.

\section{Inclusion of interaction between dark energy and CDM}
\begin{figure}
\centering
\includegraphics[width=3.3in]{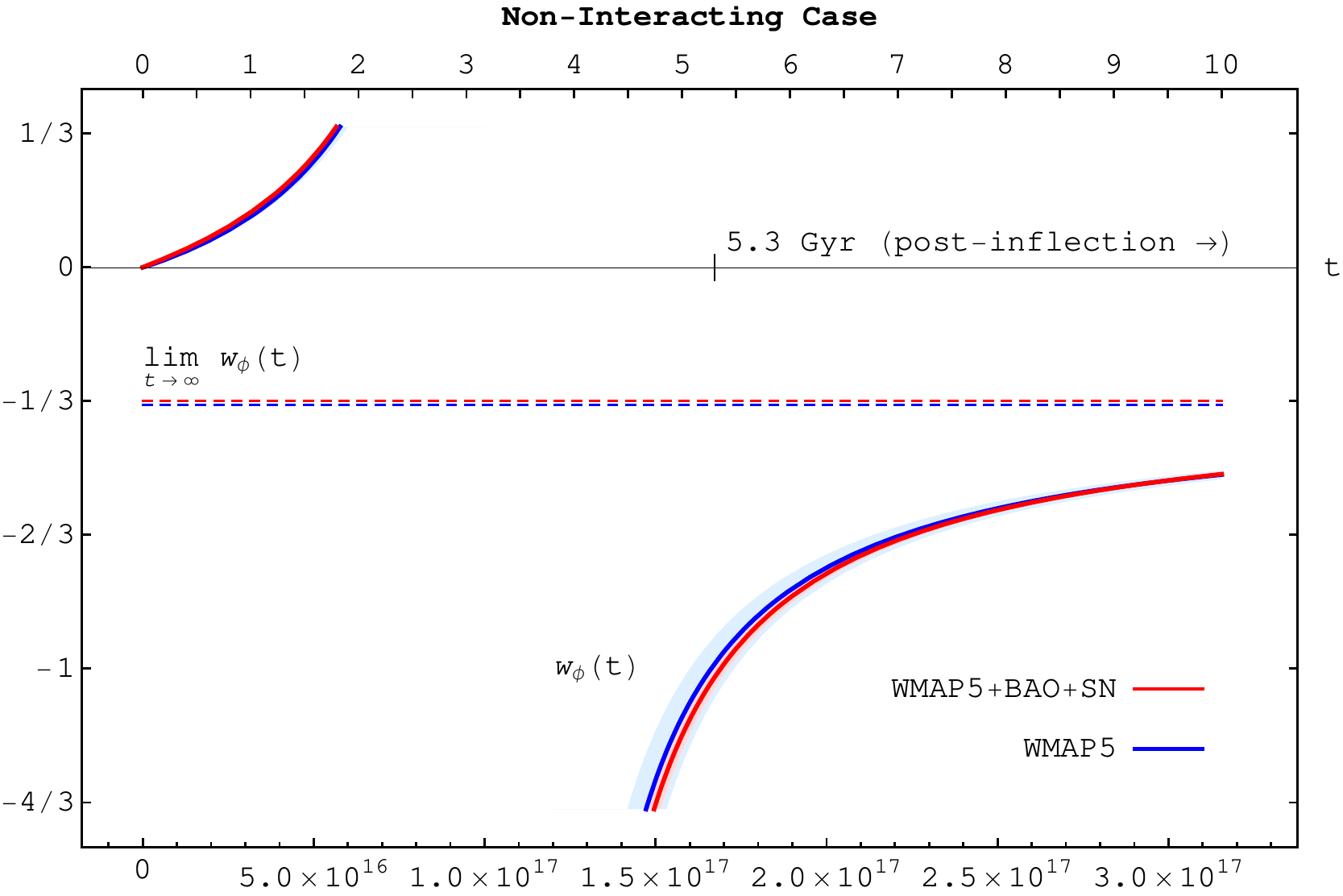}
\includegraphics[width=3.3in]{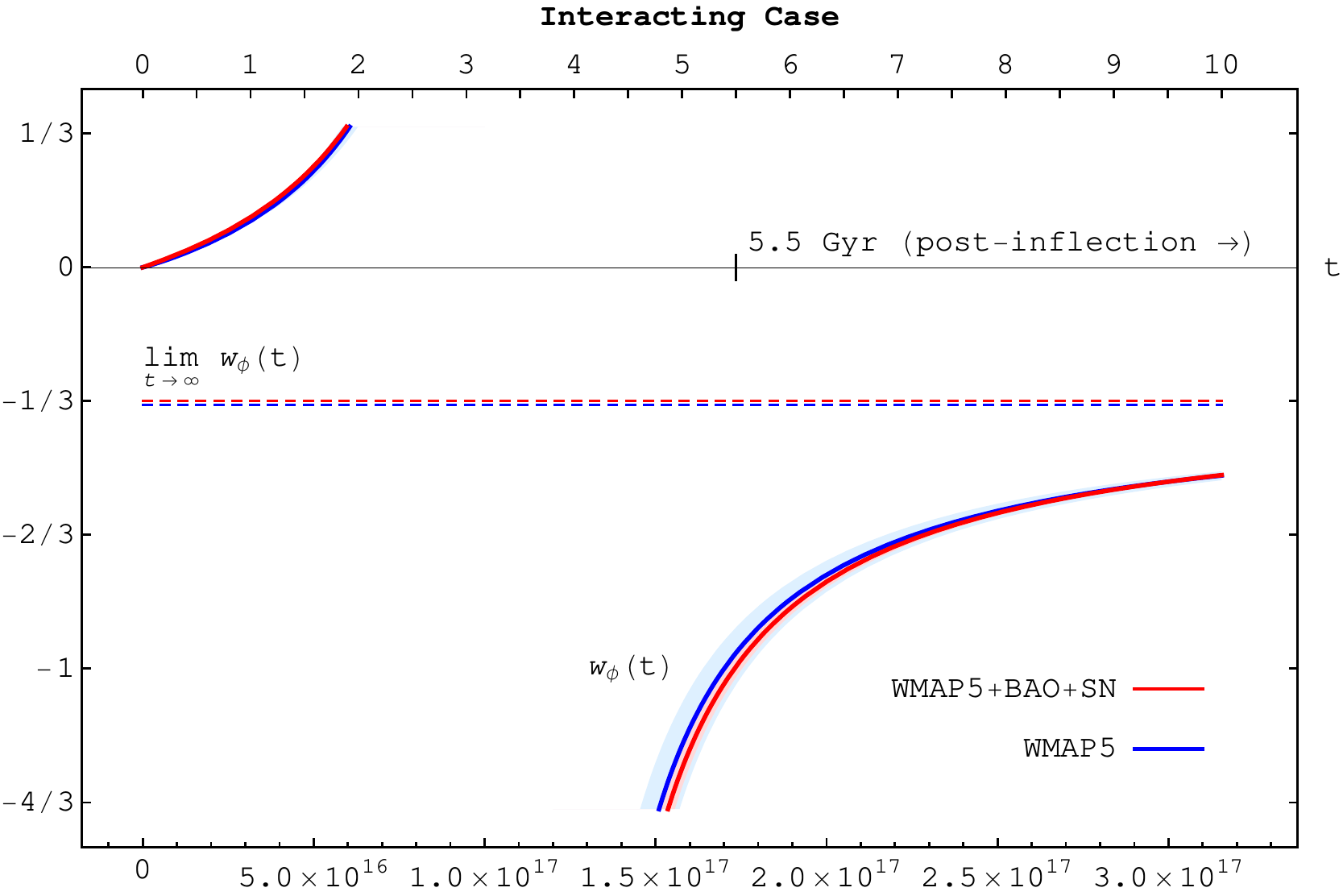}
\caption{Equation of state parameter, $w_{\phi}$ for non-interaction and interaction cases. The dash lines are asymptotic value at late time. Horizontal axis are in sec and Gyr. 
The light-shaded regions correspond to the ranges of $\Omega_{k, 0}$, and the central solid lines to $\bar\Omega_{k, 0}$. Interaction affects in slightly lowering the $w_{\phi}$ value due to decaying of dark matter to dark energy. In post-inflection period, their values (non-interaction and interaction cases) are less and less different due to the subdomination of the dust density term.}
\label{w} 
\end{figure}

\begin{figure}
\centering
\includegraphics[width=3.3in]{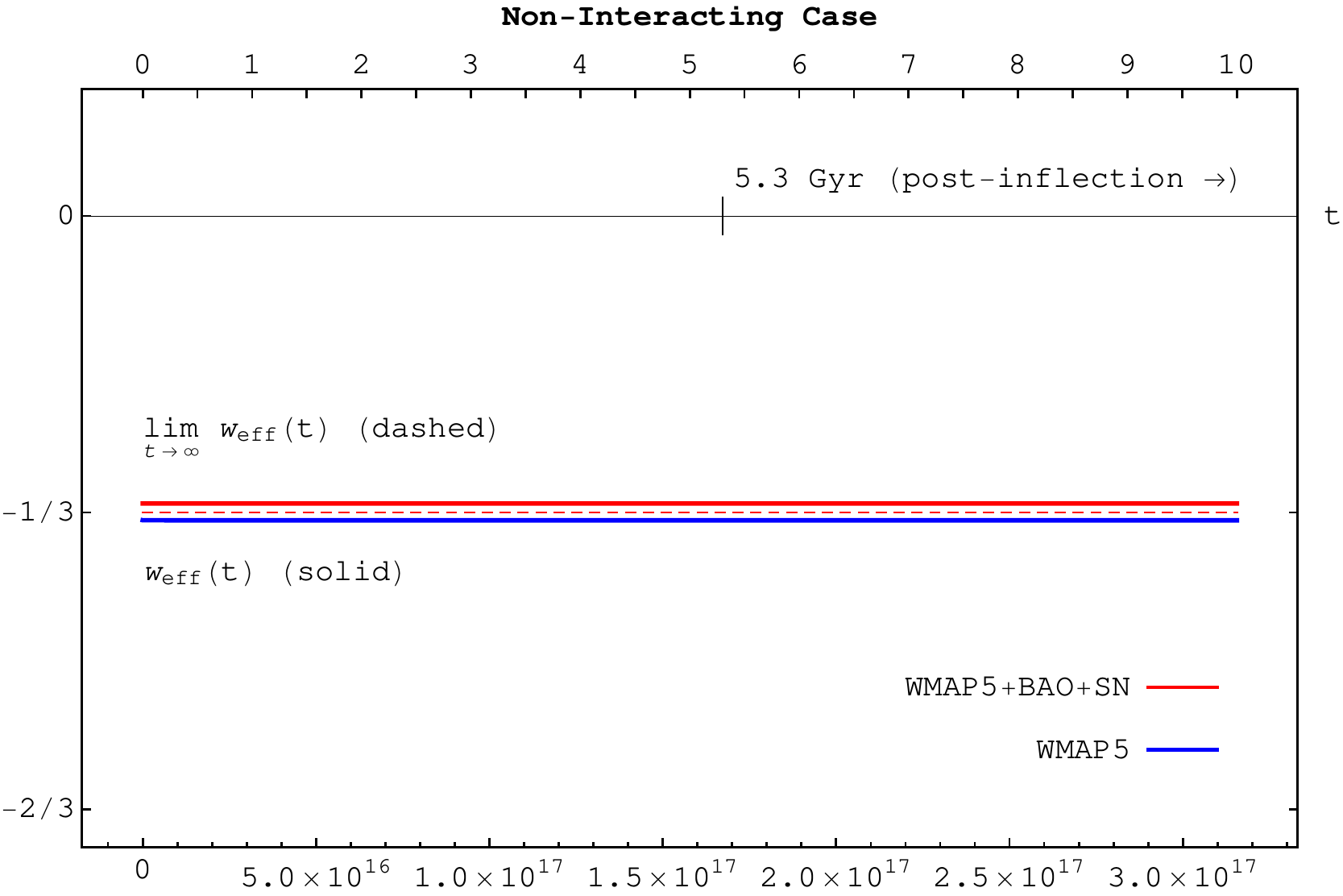}
\includegraphics[width=3.3in]{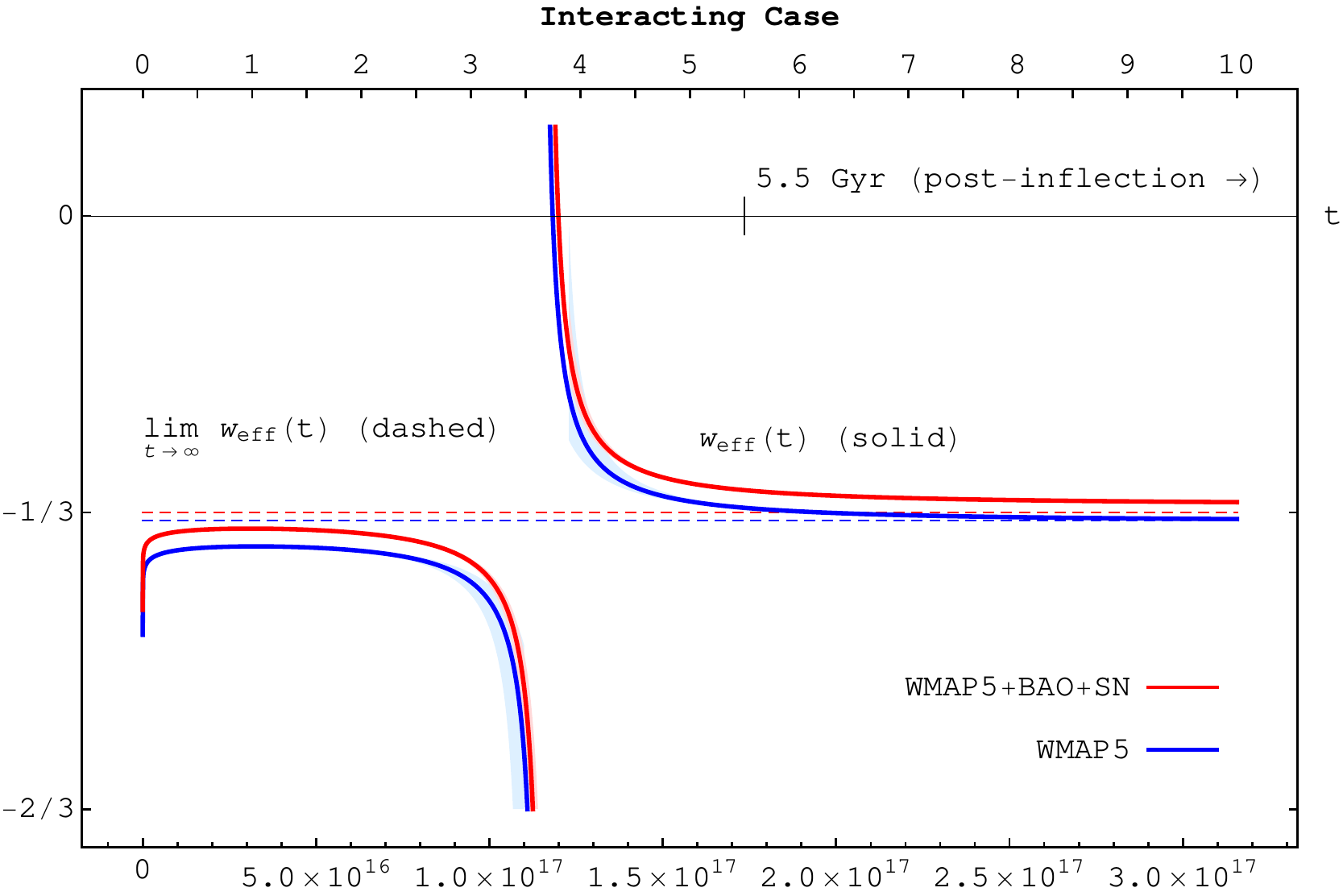}
\caption{Weighted effective equation of state parameter, $w_{\rm eff}$ for non-interaction and interaction cases. Interaction effect lifts the values of $w_{\rm eff}$ at earlier time before entering post-inflection era when interaction effect is less important due to subdomination of the dust term. In both non-interaction and interaction cases, $w_{\rm eff}$ diverges when time satisfies condition,  $t^{2(\alpha -1)} = - k ( c t_0^{\alpha} / \alpha a_0  )^2  $. Notice that here $k > 0$. Existence of divergency in $w_{\rm eff}$, (condition that $t$ is real) depends on the value of $\alpha -1$.} 
\label{wg} 
\end{figure}

\begin{figure}
\centering
\includegraphics[width=3.3in]{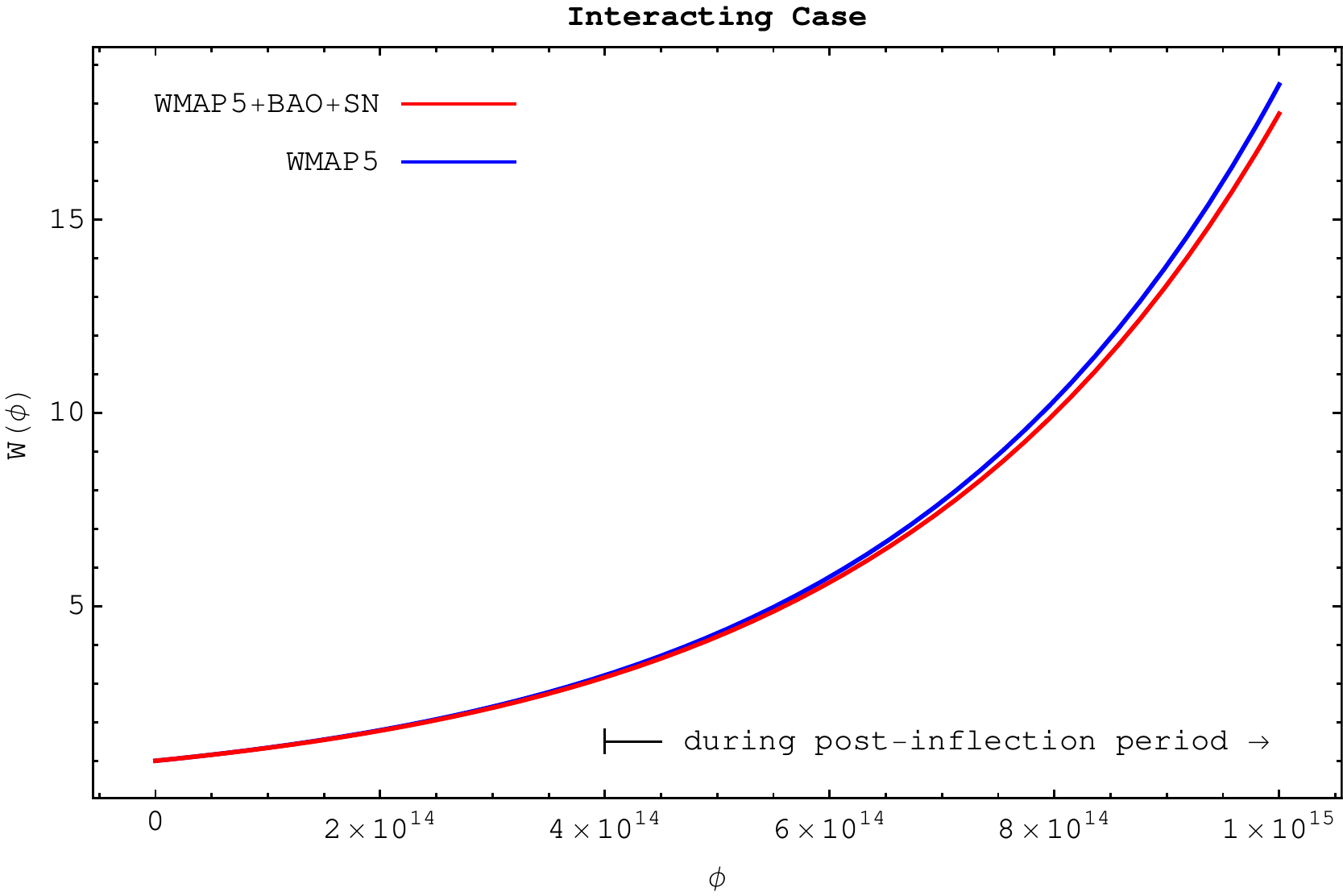}
\caption{Mass function $W(\phi)$ of the dark matter-dark energy interaction term versus $\phi$.  The mass function declines sharply at first and slowly converges to zero. $t_\textrm{inflection}$ corresponds to $\phi \approx 4 \times 10^{14}$ for both datasets.}
\label{interact-wbig}
\end{figure}

 In situation such that there is an interaction between the scalar field and cold dark matter which is a non-baryonic sector of dust \citep{Amendola:1999er},  we can have interesting scenario. This can result from adding an Yukawa-like interacting term, $-W(\phi) m_0 \bar{\psi}\psi$ to the Lagrangian density where $\psi$ is fermionic cold dark matter field  as in \citep{Das:2005yj} and \citep{Amendola:2006dg}. Other studies about dark sectors interaction can be seen in \citep{He:2008tn}    The fluid equations for cold dark matter and the scalar field are then 
\begin{eqnarray}
\dot{\rho}_{\rm CDM} + 3 H \rho_{\rm CDM} \: &=& \: \Gamma \rho_{\rm CDM}    \label{CDMint}  \,,\\
\dot{\rho}_{\phi} + 3 H \rho_{\phi}(1 + w_{\phi})  \: &=& \:  -  \Gamma \rho_{\rm CDM} \,,
\end{eqnarray} 
where the interaction rate is defined in term of the Hubble rate, $\Gamma  =  \delta(a) H$ or of the field speed $  \Gamma = Q \dot{\phi}$   
(\citep{Amendola:1999er} and \citep{Wetterich:1994bg}) and $\delta(a)$ is function of scale factor. 
This function relates to mass, $m_{\psi}$ of the  fermionic CDM as 
\begin{equation}
\delta(a)   =   \frac{{\dee} ( \ln m_{\psi}(a))
}{{\dee}  (\ln a)} \,, \end{equation} 
or that is to say
$
m_{\psi}(a) = \exp \left[ {\int \delta(a) {\dee} (\ln a)}  \right].
$
The mass function $W(\phi) = m_{\psi}(\phi(a))/m_0 $ in the Lagrangian is hence related to $\delta$ and $\Gamma$ as
\begin{equation}
\delta(a)   =   \frac{{\dee} ( {\ln} W (a))}{{\dee}  ( {\ln}  a)} \,,
\end{equation}  
and
\begin{equation}
\Gamma  \, = \, \delta(a)\, H\, =\, \frac{1}{W} \frac{{\dee} W}{{\dee} \phi} \dot{\phi} \,,
\end{equation}
where $m_0$ is the present mass of the fermion field.  
Therefore $Q = ({1}/{W})({\dee W}/{\dee \phi}) $ and this can be integrated to obtain
\begin{equation}
\delta \, = \, \frac{a}{W} \left( \frac{{\dee} W}{{\dee} a}  \right) \,,
\end{equation} 
and we found that  
\begin{equation}
\frac{W(a)}{W_0} \,= \, \left( \frac{a}{a_0}\right)^{\delta}\,,   \label{eqW}
\end{equation}
assuming constant $\delta$.  The interaction between dark matter and scalar field affects CDM matter density. It is straightforward  from (\ref{CDMint})  to find that 
\begin{equation}
\rho_{\rm CDM} \, =\,  \rho_{\rm CDM, 0} \left( \frac{a_0}{a} \right)^{3-\delta} \,.  
\end{equation}  
The equation (\ref{KG}) is also modified to
\begin{equation}
\ddot{\phi} + 3 H \dot{\phi}  \: \, =\, \:  - \, \rho_{\rm CDM} c^2 \left( \frac{1}{W} \frac{\dee W}{\dee \phi} \right) \, - \, \frac{\dee V}{\dee \phi}\,.
\end{equation}  
After integrating the first term on right hand side, the effective scalar potential is found to be
\begin{equation} 
V_{\rm eff} \;= \;    \left( \frac{\delta}{\delta-3} \right) \frac{\rho_{\rm CDM, 0} \,c^2 }{a^{3-\delta}} \,+\, V(\phi)\,.
 \end{equation}
The CDM density term  is modified due to the interaction while the baryonic matter does not interact with the scalar field (this happens in chameleon scenario \citep{Khoury:2003aq}).  Hence in this situation, the equation (\ref{phidots}) is modified to
\begin{equation} 
\dot{\phi}^2(t) = - \frac{2 M_\mathrm{P}^2 c}{\hbar} \left( \dot{H} - \frac{kc^2}{a^2} \right) - \frac{\rho_{\rm CDM, 0}c^2}{a^{3-\delta}}   -  \frac{\rho_{b, 0}c^2}{a^{3}} \,,  \label{phidotsMod} 
\end{equation}
as well as the potential
\begin{eqnarray}\label{vraw} 
V &=& \frac{3 M_{\rm P}^2 c}{\hbar } \left( H^2 + \frac{\dot{H}}{3} + \frac{2k c^2}{3a^2} \right)    -   \frac{\rho_{\rm CDM, 0} c^2}{2 a^{3-\delta}}  \nonumber \\    
 &  &  \;\;\;\;\;\;\;\;\;\;\;\;\;\;\;\;\;\;\;\;\;\;\;\;\;\;\;\;\; \;\;\;\;\;\;\;\;\;\;\;\;\;\; -   \frac{\rho_{b, 0} c^2}{2 a^{3}}. 
\end{eqnarray}
With post-inflection ap?proximation, the potential is written as function of the scalar field as,
\begin{eqnarray}
V(\phi) & \approx &  
\frac{M_{\rm P}^2 c}{\hbar} \Bigg[  \left( \frac{3\alpha^2 - \alpha}{t_0^2}  \right) e^{-2 \phi / {\sqrt{2 \alpha M_{\rm P}^2 c/ \hbar}} } \nonumber \\   &   & + \frac{2 k c^2}{a_0^2}\, e^{-2 \sqrt{\alpha}\phi / \sqrt{2 M_{\rm P}^2 c/ \hbar }}     \Bigg] 
\nonumber \\ 
&   & -  \frac{\rho_{\rm CDM, 0} \,c^2}{2 a_0^{3-\delta} }\, e^{-\sqrt{\alpha}(3-\delta)  \phi / \sqrt{2 M_{\rm P}^2 c/ \hbar }}  \nonumber \\ 
&   & -  \frac{\rho_{b, 0}\, c^2}{2 a_0^3 }\, e^{-3 \sqrt{\alpha}\phi / \sqrt{2 M_{\rm P}^2 c/ \hbar }}\,.
\end{eqnarray}
Dark energy density that decays to dark matter density is $ 
\Delta \rho_{\phi} = {\rho_{\rm CDM, 0}}/{a^3}   -   {\rho_{\rm CDM, 0}}/{a^{3-\delta}} \,.
$
For a positive $\delta$, the scalar field decays into dark matter hence $\Delta \rho_{\phi} < 0 $ while negative $\delta$ gives reverse process. We define effective scalar field density as 
$\rho_{\phi}^{\rm eff}   \equiv   \rho_{\phi}   +  \Delta \rho_{\phi}$ hence,
\begin{equation}
\rho_{\phi}^{\rm eff}     =   \rho_{\phi}   +   \frac{\rho_{\rm CDM, 0}}{a^3} \left(1 - a^{\delta} \right)\,.
\end{equation}
To find scalar field effective equation of state parameter, $w_{\phi}^{\rm eff}  $, we consider effective scalar field fluid equation,
\begin{equation}
 \frac{\dee \rho_{\phi}^{\rm eff}   }{\dee t}  \:=\:  - 3 H \rho_{\phi}^{\rm eff}   \left(1  +  w_{\phi}^{\rm eff} \right) \,. 
\end{equation}
Therefore,
$1  +  w_{\phi}^{\rm eff}   =  ({1}/{  \rho_{\phi}^{\rm eff} }) \left(   \rho_{\phi} + \Delta \rho_{\phi}  + p_{\phi} + \Delta p_{\phi}  \right)$. Note that change in scalar field density, $\Delta \rho_{\phi}$, results in change of scalar pressure density, $\Delta p_{\phi}$. With the fact that, $\Delta p_{\phi} =  w_{\phi}(\Delta \rho_{\phi}) c^2 $, therefore indeed $w_{\phi}^{\rm eff} =  w_{\phi}$, i.e. unchanged. This is because of change in density would equivalently create change in pressure (this defers from  \citep{Das:2005yj}). In order to account for total effective equation of state of all cosmic fluids, we define weighted equation of state parameter as
\begin{equation}
 w_{\rm eff}  =  \frac{p_{\phi} + w_{\phi}  (\Delta \rho_{\phi}) c^2}{\rho_{\rm CDM} c^2 + \rho_b c^2+ \rho_{\phi} c^2}\,,
\end{equation}
with the subscript eff instead of superscript. We pursue calculating all quantities as done in the non-interaction case. At present $t = t_0$, for interaction case we set $a_0=1$ with constant interaction rate $\delta = -0.03$ obtained from \citep{Guo2007}. Using WMAP5+BAO+SN combined dataset, the potential (as function of time) is
\begin{equation}\label{firstpotential_int} 
V =  \frac{1.03\cdot 10^{26}}{t^2}\, +\, \frac{1.51\cdot 10^{23}}{t^{1.97}} \,-\, \frac{2.49\cdot 10^{41}}{t^{2.95}}\, -\, \frac{4.16\cdot 10^{42}}{t^{2.98}}, \end{equation}
whereas, for WMAP5 dataset alone,
\begin{equation}\label{secondpotential_int}
 V = \frac{1.11\cdot 10^{26}}{t^2} \,+ \,\frac{7.67\cdot 10^{24}}{t^{2.03}} \,-\, \frac{8.12\cdot 10^{42}}{t^{3.04}}\, -\, \frac{1.33\cdot 10^{44}}{t^{3.07}},
 \end{equation}
 in SI unit. The field solution for WMAP5+BAO+SN combined data is 
\begin{eqnarray}
\phi(t) &=& \int_t^{t_0} \Bigg{(}\frac{1.06\cdot 10^{26}}{{\tilde t}^2}  + \frac{1.51\cdot 10^{23}}{{\tilde t}^{1.97}} - \frac{4.98\cdot 10^{41}}{{\tilde t}^{2.95}} 
 \nonumber \\ 
  & &  \;\;\;\;\;\;\;\; \;\;\;\;\;\;\;\;\;\;\;\;\;\;\;\;   - \frac{8.33 \cdot 10^{42}}{{\tilde t}^{2.98}}\Bigg{)}^{1/2}  \dee {\tilde t}\,, \\
\phi(t)  &\approx&  4.17\E{14} - 1.03\E{13} \ln t \,,  \label{eq_solint1}
\end{eqnarray}
and for the WMAP5 alone, 
\begin{eqnarray}
\phi(t) & = & \int_t^{t_0}  \Bigg{(}\frac{1.09\cdot 10^{26}}{{\tilde t}^2} + \frac{7.67\cdot 10^{24}}{{\tilde t}^{2.03}} - \frac{1.63\cdot 10^{43}}{{\tilde t}^{3.04}}    
\nonumber \\    &   &   \;\;\;\;\;\;\;\; \;\;\;\;\;\;\;\;\;\;\;\;\;\;\;\;   - \frac{2.66\cdot 10^{44}}{{\tilde t}^{3.07} }   \Bigg{)}^{1/2}  \dee {\tilde t}\,,      \\
\phi(t) & \approx& 4.23\E{14} - 1.04\E{13} \ln t \,,     \label{eq_solint2}
\end{eqnarray}
with estimation that the barotropic density and spatial curvature term are subdominant as before.  Using  (\ref{eq_solint1}) with WMAP5+BAO+SN dataset, the scalar potential is 
\begin{align}
V(\phi) \approx & {2.72\cdot 10^{-12}}{e^{1.92\cdot 10^{-13} \phi}}  \nonumber \\
           & + {5.51\cdot 10^{-10}}{e^{1.95\cdot 10^{-13} \phi}} \nonumber \\
            & - {1.91\cdot 10^{-11}}{e^{2.88\cdot 10^{-13} \phi}} \nonumber \\
           & - {9.61\cdot 10^{-11}}{e^{2.90\cdot 10^{-13} \phi}}\,.
\end{align}
and using (\ref{eq_solint2}) with WMAP5 alone, the potential is 
\begin{align}
V(\phi) \approx & {5.94\cdot 10^{-10}}{e^{1.92\cdot 10^{-13} \phi}}   \nonumber \\
           &  + {1.36\cdot 10^{-11}}{e^{1.94\cdot 10^{-13} \phi}}   \nonumber \\
           & - {1.91\cdot 10^{-11}}{e^{2.92\cdot 10^{-13} \phi}}  \nonumber \\
           &  - {9.12\cdot 10^{-11}}{e^{2.95\cdot 10^{-13} \phi}}\,.
\end{align}
The equation of state for the scalar field and weighted value with dark matter interaction as function of time can be found directly as previously. These are in form of complicated functions plotted in Figs. \ref{w} and \ref{wg}. The values  at present are
\begin{equation}
w_\phi (t_0) = - 0.446\,,\;\;\;\;   w_{\rm eff}  (t_0) = - 0.323\,,
\end{equation}
 for  WMAP5+BAO+SN data  and for the WMAP5 dataset alone they are 
\begin{equation}
w_\phi (t_0) = - 0.452\,,\;\;\;\;  w_{\rm eff} (t_0) = - 0.342\,.
\end{equation}
For both datasets, $w_\phi (t)$ and $w_{\rm eff} (t)$ converge to approximately $-1/3$. 
In post-inflection period, non-interaction and interaction cases are less and less different due to the subdomination of the dust density term.
These values at present time for non-interaction and interaction cases are the same at three significant decimal digits.
The value of $w_{\phi}$ diverges when time satisfies the condition, 
\begin{equation}
\alpha^2 = \frac{\hbar  \, c}{3 M_{\rm P}^2 }  \left(  \frac{\rho_{\rm CDM, 0} \,  t_0^{-\alpha \delta} }{a_0^{-\delta} \, t^{- \alpha \delta}} + \rho_{b, 0}   \right) \frac{t_0^{3 \alpha}}{a_0^3  t^{3 \alpha}}-   \frac{k c^2 \, t_0^{2 \alpha}}{a_0^2  \, t^{2 \alpha - 2}}\,.
\end{equation}
and for $w_{\rm eff}$, it diverges when time satisfies condition,  $t^{2(\alpha -1)} = - k ( c t_0^{\alpha} / \alpha a_0  )^2  $ as in non-interaction case. 
Considering mass potential function  (\ref{eqW}) in scalar-field power-law cosmology, it is simply
$ {W}  = W_0 \left( {t}/{t_0}   \right)^{\alpha \delta} $. With similar procedure as used in finding $V(\phi)$, we obtain 
\begin{align}
W(\phi) &\approx  W_0 \, \exp{\left[\left( \frac{\sqrt{\alpha} \,\delta}{\sqrt{2 M_{\rm P}^2 c / \hbar}} \right)\phi \right]}  \,,   \\
Q &\approx     \frac{\sqrt{\alpha} \,\delta}{\sqrt{2 M_{\rm P}^2 c / \hbar}}\,.
\end{align}
The mass potential function is exponentially increasing if $\delta > 0$, i.e. when dark energy decays into dark matter. On the other hand, in our case, $\delta = -0.03$, we have exponentially decay mass potential function. The result for both datasets are
\begin{align}
W(\phi) &\approx e^{2.88\E{-15} \phi} \,,   \\
Q &\approx  2.88\E{-15} \,, 
\end{align}
for  WMAP5+BAO+SN data and
\begin{align}
W(\phi) &\approx e^{2.92\E{-15} \phi} \,,   \\
Q &\approx 2.92\E{-15} \,,  
\end{align}
using WMAP5 data alone. We set $W_0 = W(\phi(a_0)) = 1$.  $W(\phi)$ is plotted in Fig. \ref{interact-wbig}. 

\section{Conclusion}
We consider scalar-field power-law cosmology featuring homogeneous canonical scalar field and dust fluid evolving in a late non-flat FLRW universe expanding with power-law function.  The scalar field is minimally coupled to gravity and dust fluids are baryonic and CDM perfect fluids. We use two sets of observational data, combined WMAP5+BAO+SN dataset and WMAP5 dataset, as the inputs.   Mean values of both sets suggest slightly closed geometry. With $k = +1$, it gives mean value of $a_0$ at $8.4\E{26}$ for WMAP5 and $1.9\E{27}$ for WMAP5+BAO+SN datasets. For closed universe, the WMAP5 dataset puts the lower limit of $5.1\E{26}$ for $a_0$ while the WMAP5+BAO+SN dataset dataset puts the lower limit of $9.85\E{26}$. The WMAP5 data alone yields the exponent $\alpha = 1.01$ agreeing with the previous study of $H(z)$ data which gives $\alpha = 1.07$ \citep{Dev:2008ey}. However combined WMAP5+BAO+SN data disagrees with the previous combined $H(z)+$SN Ia data, i.e. WMAP5+BAO+SN gives $\alpha=0.985$ while $H(z)$+SN gives $\alpha = 2.28$ (closed geometry)  \citep{Dev:2008ey}. 

The scalar field potential is found with the two datasets. Slope of the potential is inflected after $t_{\rm inflection}$. This time scale characterizes the time when the Hubble expansion and $\dot{H}$ terms begin to dominate over the dust and curvature terms. For the WMAP5 dataset,  $t_{\rm inflection}=$5.29 Gyr and for the WMAP5+BAO+SN dataset,  $t_{\rm inflection}=$ 5.33 Gyr. We also consider situation when there is interaction between dark energy-dark matter in order to see its effects on the potential and equation of state. Constant interaction rate is assumed here with $\delta = -0.03$ from \citep{Guo2007} corresponding to dark matter decaying into the scalar field. For interaction case, $t_{\rm inflection}=$5.48 Gyr (WMAP5 dataset) and $t_{\rm inflection}=$ 5.55 Gyr (WMAP5+BAO+SN dataset). The interaction affects in lowering the height of scalar potential and shifting the potential rightwards to later time (see Fig. \ref{potentialplots} and Table \ref{resulttable}). In this study, with closed geometry suggested by WMAP5 data, the field equation of state does not match the WMAP5 observation ($w$CDM model), i.e. $w \approx -1$.
Effects of interaction is in slightly lowering the $w_{\phi}$ value while slightly lifting up the weighted effective equation of state $w_{\rm eff}$ at earlier time. We found that the mass potential function $W(\phi)$ of the interaction is approximately in form of exponential decay function.

\section*{Acknowledgments}

We thank Chris Clarkson for discussion. The authors thank to the referee for useful comments. K.~T. is supported by a research assistantship under a grant of the Thailand Toray Science Foundation (TTSF) and a NARIT postgraduate studentship. B.~G. is sponsored by the Thailand Research Fund and the National Research Council of Thailand.


\end{document}